\documentclass[aps,pre,10pt,twocolumn,superscriptaddress,nofootinbib,floatfix,hidelinks]{revtex4-2}
\usepackage{newtxtext}
\usepackage{amsmath,amssymb}
\usepackage{bm}
\usepackage{graphicx}
\usepackage[colorlinks = true, linkcolor = blue, citecolor = blue]{hyperref}

\newcommand{\D}{\mathcal{D}}

\begin{document}
\setlength{\parskip}{0pt}

\title{Keeping the interaction structure explicit:\texorpdfstring{\\ }{ }comment on ``Graphs are maximally expressive for higher-order interactions''}

\author{Giulio Burgio}
\affiliation{Institute of Functional Biology and Genomics IBFG-CSIC, University of Salamanca, 37007 Salamanca, Spain}

\begin{abstract}
Peixoto \emph{et al.}~\cite{peixoto2026graphs} stress that graph-based formulations can express any interaction model, a point the literature on higher-order networks (HONs) has often overlooked. We agree with much of their critique. We argue that expressiveness, however, might not be the main reason hypergraphs are often preferred. A network is read from a model rather than assumed, and to properly study the role of structure---one of the most basic questions in complex systems research---one needs a representation that keeps that structure separate from the functional form of the interactions. We show that such representation is in general a directed hypergraph (equivalently, a directed factor graph), with one hyperarc per interaction term, recovering a (directed) graph for pairwise interactions, and an undirected hypergraph for symmetric, higher-order ones. Most of the HONs literature has focused so far on symmetric interactions, which, in light of what is established here, might explain why there (undirected) hypergraphs are regularly presented as the natural representation for systems with higher-order interactions.
\end{abstract}

\maketitle

An interaction model coupling the variables of a system always admits a specific graph representation that is unique and not arbitrary. Such representation, commonly known as \emph{dependency graph} in computer science and statistics, has a vertex for each of the $N$ variables and a directed edge from vertex $j$ to vertex $i$ if the variable at $i$ depends directly on the variable at $j$. The dependency graph is thus not assumed but derived from the model. Just to fix the concepts, let us consider from now on the class of systems described by ODEs,
\begin{equation}
    \dot x_i=f_i(\boldsymbol x)\; ,
    \label{system}
\end{equation}
for $i = 1,\dots,N$. The dependency graph $G$ associated to Eq.~(\ref{system}) contains the edge $j\to i$ if $f_i$ depends on $x_j$ (which can be operationalized via the condition $\partial f_i/\partial x_j\not\equiv0$)~\cite{peixoto2026graphs}. Denoting with $\bm W$ the adjacency matrix of $G$, $W_{ij} = 1$ if $j\to i$ exists and $W_{ij} = 0$ otherwise. The change of $x_i$ is fully determined by the variables $\bm{x}_{\partial i} = \{x_j: j\in\partial i\}$ in the in-neighborhood $\partial i = \{j: W_{ij}=1\}$ of $i$, so that Eq.~(\ref{system}) can be equivalently expressed as\footnote{Notice that self-interaction terms (i.e., univariate terms $g(x_i)$ in $f_i$), if any, are also accounted for; $\partial i$ contains $i$ itself in this case.}
\begin{equation}
    \dot x_i=f_i(\bm{x}_{\partial i})\; .
    \label{system2}
\end{equation}
Equation~(\ref{system2}) can express any interaction model; and the dependency graph, being a derived object, plays no role in this~\cite{peixoto2026graphs}.

What is interesting about Eq.~(\ref{system2}) is not its trivial maximal expressivity, but the network analysis one can make out of the dependency graph it induces. A fundamental question in network science is to understand the role the structure of dependencies plays in determining the properties of a model (or a family of models)~\cite{barrat2008dynamical,barabasi2016network,newman2018networks}. Defining an interaction model amounts to specifying (i) each dependency among variables and (ii) the functional form of each dependency. Understanding the role of the structure of dependencies thus means isolating as much as possible the effect of (i) from that of (ii).\footnote{Models where variables and structure coevolve of course blur the line between (i) and (ii). Yet, generally, there are structural properties that are conserved and thus worth studying on their own~\cite{gross2009adaptive,burgio2025characteristic}.}

When each interaction involves at most two variables, the general model reads (Eq.~(18) of Ref.~\cite{peixoto2026graphs})
\begin{equation}
    f_i(\bm{x}_{\partial i}) = \sum_{j\in\partial i} h_{i,j}(x_i,x_j) = \sum_{j} W_{ij} h_{i,j}(x_i,x_j) \; .
    \label{f_pair}
\end{equation}
In this case, the dependency graph $\bm W$ determines the structure of dependencies completely. That is, if we were given only $\bm W$, we could recover all and only the interactions in the model (except, of course, the functional form $h$ of each of them). This is the ``only kind of dynamics that can be unambiguously considered \emph{pairwise}''~\cite{peixoto2026graphs}, and the formal reason for this is that its structure of dependencies can be fully stored in a graph---the dependency graph. Being able to factorize the dependency structure of a model from the functional form of each dependency is what allows to address a basic question like ``what is the effect of the network?'' precisely.

% To formalize this fact, let us henceforth call a network representation of an interaction model \emph{complete} if the structure of dependencies of the model can be unequivocally recovered from it. For a pairwise model (Eq.~(\ref{f_pair})), the complete representation is the dependency graph.

To formalize this fact, let us henceforth call a network representation of an interaction model \emph{complete} if the structure of dependencies of the model can be unequivocally recovered from it.
The dependency graph always provides the complete representation for a pairwise model (Eq.~(\ref{f_pair})). It does not, however, for models including (higher-order) interactions among three or more variables (e.g., $f_i=x_j+x_k$ and $f_i=x_jx_k$ give the same neighborhood $\partial i$). It is natural to ask what is the complete representation of a general model with interactions involving arbitrary numbers of variables. To this end, we need a stronger definition of dependency.

\begin{figure}
    \centering
    \includegraphics[width=\linewidth]{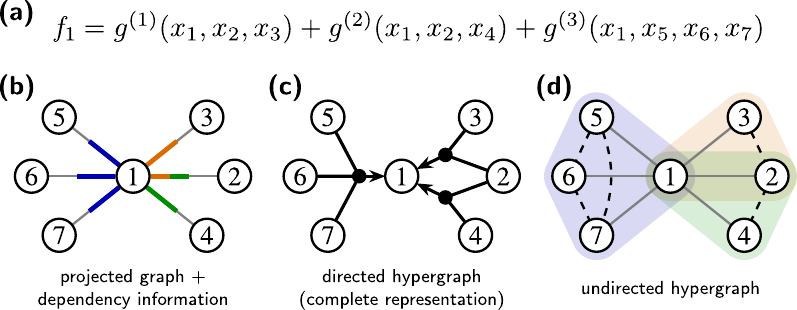}
    \caption{A toy example. (a) The model for vertex 1, consisting of three independent terms. (b) The dependency graph records the neighborhood of vertex $1$, namely which vertices affect it, but not the specific combinations in which they do so. The graph must be supplemented with such information (colored stubs, as in Fig.~3(b) in Ref.~\cite{peixoto2026graphs}) in order to account for the structure of the model. (c) The complete representation given by the directed hypergraph, with hyperarcs $\{2,3\}\to\{1\}$, $\{2,4\}\to\{1\}$ and $\{5,6,7\}\to\{1\}$ (self-interactions implicit in the tails). (d) The toy model has no cliques, so an undirected hypergraph cannot represent it without asserting spurious dependencies (dashed). It is worth noticing that making the edges in (b) directed does not resolve the structural deficiency of the dependency graph.}
    \label{fig1}
\end{figure}

%(e.g., $f_i=x_j+x_k$ implicitly assumes $r=0$)
Given a model, as specified by $\{f_i\}_i$, let us call $D$ a dependency set for vertex $i$ if $f_i$ contains a non-null term, $h_{i,D}\not\equiv0$, involving all and only the variables in $D$ (potentially including $x_i$ itself).\footnote{Terms involving the same set of variables can always be combined into a single one.} To operationalize this, let $r$ be a reference value such that each term $h_{i,D}$ in $f_i$ vanishes whenever any of the variables in $D$ equals $r$.\footnote{Not all models admit such a reference value. For example, when interactions depend exclusively on variable differences (i.e., having the global symmetry $x\to x+c$), no interaction term can vanish whenever one of its variables takes a fixed value, unless it vanishes identically. In such cases there might be no operation (such as Eq.~(\ref{h_ie})) able to retrieve the dependency sets, yet one can still read them off from the specified $f_i$. Also notice that, contrary to maximal dependency sets (i.e., those not contained in a larger one), non-maximal sets are generally not invariant under a change of variables. For instance, $x_1x_2-x_1-x_2$ ($r=0$ is the reference) becomes $y_1y_2-1$ under $x_i\to \varphi(x_i)\equiv y_i=x_i-1$ ($r = 1$).
Nonetheless, all dependency sets (whether maximal or not) are uniquely identified within any specification of the model.} Then $f_i$ can be written uniquely as a sum of terms $h_{i,D}(\boldsymbol x_D)$, one for each dependency set $D$, given by inclusion--exclusion,\footnote{The sole, weak requirement here is that each variable can take the reference value independently of the others. Also, notice that the empty dependency set $D=\emptyset$ gives the constant term $h_{i,\emptyset}=f_i(\boldsymbol r)$.}
\begin{equation}
h_{i,D}(\boldsymbol x_D)=\sum_{S\subseteq D}(-1)^{|D\setminus S|}f_i(\boldsymbol x_S,\boldsymbol r_{\setminus S}) \; ,
\label{h_ie}
\end{equation}
where $(\boldsymbol x_S,\boldsymbol r_{\setminus S})$ is the state obtained from $\boldsymbol x$ by setting all variables outside $S$ to the reference value. For example, for $f$ a polynomial and $r=0$, the dependency sets are simply the sets of variables appearing in its monomials. The set $\D_i$ of all dependency sets in $f_i$ is the complete representation of the dependency structure of vertex $i$; and therefore
\begin{equation}
f_i(\boldsymbol x_{\D_i})=\sum_{D\in\D_i}h_{i,D}(\boldsymbol x_D) \; .
\label{f_gen}
\end{equation}
Clearly, $\partial i = \bigcup_{D\in\D_i} D$, so that $\partial i$ provides a compatible but not complete representation of $f_i$. Equation~(\ref{f_gen}) makes it obvious---perhaps even more than Eq.~(20) of Ref.~\cite{peixoto2026graphs} already does---that the complete representation of a general model is a \emph{directed hypergraph}~\cite{gallo1993directed} (see also Fig.~\ref{fig1}).\footnote{Analogous observations have been made by Moutuou~\cite{moutuou2026graphs} and very recently by Rommens \emph{et al.}~\cite{rommens2026beyond}. Moutuou introduced the concept of focal hypergraph, i.e., a hypergraph endowed with a function assigning to each hyperedge one or more focal vertices (the affected ones). A focal vertex is an element of the head of a hyperarc, so that a focal hypergraph is no different from a directed hypergraph, this being a well-established concept~\cite{gallo1993directed}.} Exactly as $j\in \partial i$ translates unequivocally to the edge $j\to i$ in a pairwise model (Eq.~(\ref{f_pair})), $D \in \D_i$ translates to the hyperarc $e: D \to \{i\}$ of tail $T_e=D$ (the input variables) and head $H_e=\{i\}$ (the dependent variables).\footnote{Ref.~\cite{peixoto2026graphs} states that the notion of asymmetric interactions ``is not as easily translated for hypergraphs, since this requires making arbitrary and more restricted decisions about what constitutes an ordering of a group''. Our simple construction shows how easy it is to map any model's structure to the tails and heads of the hyperarcs of a directed hypergraph.
Moreover, the additional difficulty that interaction functions with ordered arguments (e.g., $h(x_j,x_k)\neq h(x_k,x_j)$) would pose is not specific to directed hypergraphs; dependency graphs do not record that information either.} Introducing the head incidence matrix $\bm I$ of elements $I_{ie}=1$ if $i\in H_e$ and $0$ otherwise, we can finally rewrite Eq.~(\ref{f_gen}) as
\begin{equation}
f_i(\boldsymbol x_{\D_i}) = \sum_{\substack{e:\\ i \in H_e}} h_{i,T_e}(\bm{x}_{T_e}) = \sum_{e} I_{ie}\, h_{i,T_e}(\bm{x}_{T_e}) \; ,
\label{general_dh}
\end{equation}
with $\D_i = \{T_e: i\in H_e\}$. In other words, the full factorization of a generic interaction model in (i) a structural component---the complete dependency structure of each variable---and (ii) a functional one---the functional form of each dependency---is always achieved by a directed hypergraph representation.

The complete representation reduces to common network structures for wide classes of models. In pairwise models (Eq.~(\ref{f_pair})), the complete representation is a directed graph. In models with node-centered dynamics (Eq.~(\ref{system2})), $\D$ consists of a single set covering the entire neighborhood of a node. In models with symmetric group interactions (Eq.~(19) of Ref.~\cite{peixoto2026graphs}), the complete representation is an undirected hypergraph. Separable models with interactions that are neither pairwise nor symmetric yield instead generic directed hypergraphs.

The choice among network representations is, nonetheless, ``inherently empirical and context dependent, and should ultimately be guided by evidence about which representation captures the relevant structure of the system most effectively''~\cite{peixoto2026graphs}. We agree---it depends on the question we want to make. At least when the goal is isolating the role of the interaction structure, the complete structural representation (generally a directed hypergraph) is the ideal choice.

As an example, consider the problem of how the relative distribution between interactions of different orders across a system affects a certain dynamics~\cite{zhang2023higher,burgio2024triadic,malizia2025hyperedge}. The most precise way to address it is to change that relative distribution while keeping everything else in the model untouched. We should thus fix the dynamics, namely the parameters and functional forms of the interactions, as well as any other structural property we can, for instance, the average number of interactions per vertex at each order.\footnote{This is readily implemented as follows. Associate a pairwise interaction to each edge of a graph generated with a desired average degree, this translating to the average number of pairwise interactions per vertex. Then, in a biased way, add 3-edges until the target average number of three-way interactions per vertex is reached. Add each 3-edge preferentially to subsets of vertices (not) forming a 3-clique in the original graph to get a positive (negative) inter-order correlation. Repeat (add $k$-edges) for higher orders, if any. The procedure is easily extended to asymmetric interactions.} This is what was done in Ref.~\cite{burgio2024triadic}, where we asked how macroscopic properties of a contagion process are affected by the interaction overlap (inter-order correlation) between pairwise and three-way infection channels, finding the overlap to control both the invasion threshold and the outbreak size. Since we assumed all interactions were symmetric, it felt natural formalizing the problem through undirected hypergraphs. The reason is simple: we needed to fully isolate structure from dynamics, and we achieved it using (what we here called) the complete representation of the model's dependency structure.

Moving from purely theoretical analysis to inference from data, support in favor of one description over another should be ultimately based on objective model selection criteria. Section~VII of Ref.~\cite{peixoto2026graphs} compares the minimum description length (a measure of plausibility of a model based on its ability to compress data) of graph-based and (undirected) hypergraph-based models. Repeating the analysis including directed hypergraphs leads to similar conclusions. For the model contribution to the description length, a graph description can be more or less compressive. This depends on how the interaction functions ($h$) are sampled and on how much the dependency sets (i.e., the hyperarcs of the complete representation) overlap.\footnote{Following Sec.~VII of Ref.~\cite{peixoto2026graphs}, consider discrete functions $X^k\to X$ with $|X|=A$, whose number is $\Omega(A,k)=A^{A^k}$. With uniform encoding, a hyperarc with $k_e$ vertices in the tail (head included) costs $A^{k_e}\log_2A$ bits. For a structurally sparse model where dependency sets share at most one vertex, $\sum_i\sum_{e\to i}A^{k_e}\le\sum_iA^{1+\sum_{e\to i}(k_e-1)}$, the right-hand side being the cost of the projected graph. The equality holds only if every vertex has either a single incoming hyperarc (the graph of Eq.~(\ref{system2})) or, for $A=2$, two incoming hyperarcs of size two. When dependency sets overlap, the inequality can be reversed by redundancy; e.g., for $A=2$, a vertex whose dependency sets consist of itself together with any two of its four neighbors costs 48 entries against the graph's 32.} Nonetheless, unlike undirected ones, directed hypergraphs are always admissible.

\paragraph*{Final remarks.} The critique put forward by Peixoto \emph{et al.}~\cite{peixoto2026graphs} of the HONs literature~\cite{battiston2020networks,torres2021and,battiston2021physics,majhi2022dynamics,battiston2022higher,bick2023higher,ferraz2024contagion,battiston2026collective} (and references therein) is well justified. We agree that this literature should have been positioned more cautiously and transparently within the broad body of work in complex systems. At the same time, we have no doubts about the relevance that hypergraph-based representations can have in studying systems with higher-order (multivariate) dependencies~\cite{lacasa2026equivalence,arenas2026higher,rommens2026beyond}. Which representation to choose depends on the problem at hand. The one-to-one mapping between a generic interaction model and its complete network representation as a directed hypergraph (or factor graph) helps in the choice. When the question is how the structural organization of interactions shapes the behavior of a system---one of the fundamental questions in complex systems research---, the complete representation is the one making this organization fully explicit. Most of the HONs literature has (understandably) so far focused on symmetric interactions, for which undirected hypergraphs offer, in fact, the complete representation. Accordingly, it might not be a coincidence that HONs studies often present (undirected) hypergraphs as the most ``natural'' or ``direct'' representation for systems with higher-order interactions.\footnote{It might also be that most people find hypergraphs a more intuitive mental model than equivalent factor graphs. Science is human after all.}

\paragraph*{Acknowledgements.} The author is grateful to Laurent H\'ebert-Dufresne for valuable feedback that improved the clarity and presentation of this work.

\bibliographystyle{apsrev4-2}
\bibliography{biblio}

\end{document}